\documentstyle[12pt,psfig,amssymb]{article}        
\begin{document}
\baselineskip=24pt
\bibliographystyle{unsrt}

\newcommand{\bmu}{{\bf \mu}}

\begin{center}
 
{\Large{\bf 
Dynamics of Bound and Free Water in an Aqueous Micellar Solution : Analysis of
the Lifetime and Vibrational Frequencies of Hydrogen Bonds at a Complex 
Interface
}}\\
\vspace{1cm}
{\large{\bf Subrata Pal$^{b}$,  Sundaram Balasubramanian$^{a}$\footnote{email:bala@jncasr.ac.in}$^{*}$,   
and Biman Bagchi$^{b}$\footnote{email:bbagchi@sscu.iisc.ernet.in}$^{*}$}}\\
\vspace{0.1cm}
$^{a}$ Chemistry and Physics of Materials Unit, \\
Jawaharlal Nehru Centre for Advanced Scientific Research,\\
Jakkur, Bangalore 560064, India.\\
$^{b}$Solid State and Structural Chemistry Unit,
Indian  Institute of Science,\\
Bangalore 560012, India.\\ 
\end{center}
\begin{center}
{\large {\bf Abstract} 
}
\end{center} 
In order to understand the nature and dynamics of interfacial water molecules
on the surface of complex systems, large scale, fully atomistic molecular
dynamics simulations of an aqueous micelle of cesium perfluorooctanoate (CsPFO)
surfactant molecules have been carried out.  The lifetime and the
intermolecular vibrational frequencies of the hydrogen bonds that the water
molecules form with the hydrophilic, polar headgroups (PHG) of the surfactants,
are calculated.  Our earlier classification~\cite{cs02,bala_condmat,pal_jpcb} of the
interfacial water molecules, based on structural and energetic considerations,
into bound and free type is further validated by their dynamics.  Lifetime
correlation functions of the water-surfactant hydrogen bonds show the long
lived nature of the bound water species. Surprisingly, the water molecules that
are singly hydrogen bonded to the surfactants have longer lifetime than  those
that form two such hydrogen bonds.  The free water molecules that do not form
any such hydrogen bonds behave similar to bulk water in their reorientational
dynamics.  
A few water molecules that form two such hydrogen bonds are
orientationally locked in for durations of the order of few hundreds of
picoseconds, that is, much longer than their {\it average} lifetime.  The
intermolecular vibrational frequencies of these interfacial water molecules
have been studied from the power spectra of their velocity autocorrelation
function.  We find a significant blue shift in the librational band of the
interfacial water molecules, apart from a similar shift in the near neighbor
bending modes, relative to water molecules in bulk. These blue shifts suggest
an increase in rigidity in the structure around interfacial water molecules.
This is in good agreement with recent incoherent, inelastic neutron scattering
(IINS) data on macromolecular solutions~\cite{ruffle}. The results of the
present simulations appear to be rather general and should be relevant to the
understanding of dynamics of water near any hydrophilic surface.

\newpage

\section{Introduction}
Water molecules which are present on the surface of soft, complex systems have
been the focus of attention for a long
time~\cite{grant_gaiduk,lee_rossky,fleming95,fourkas02,rossky98,lang99,tarek_tobias,nandi00}.
Not only are the intrinsic structure and dynamics of such water molecules
themselves interesting, but their ability to influence the structure, dynamics
and function (in the case of biologically relevant complex systems) has made
the study of interfacial water a subject of paramount interest to a wide spectrum of
researchers from different areas~\cite{biopapers}. A variety of experimental,
computational, and theoretical tools have been employed to study its
properties. Dielectric relaxation experiments have shown the presence of a
distinct relaxation peak that could be related to the dynamics of water bound
to the surface of a macromolecule~\cite{buchner,steinhauser}. Evidence for the
existence of bound water, and its slow dynamics has also been provided by local
probes such as NMR~\cite{halle99}, and solvation dynamics
experiments~\cite{zewailfeature,sarkar96,riter98,kankan02}. 

Computer simulations have enormously aided our understanding of these complex
systems~\cite{sanjoy98,rocchi,ladanyi,bala1,bala2,paljcp,currsci,balaprl,tarek_tobias}.
Employing state of the art molecular dynamics simulations, we have recently
established the presence of a slow component, running into hundreds of
picoseconds, both in the orientational relaxation of the interfacial water molecules
and also in the solvation dynamics of ions that are present near the
water-micelle interface~\cite{bala1,bala2,currsci}, consistent with time resolved
fluorescence experiments~\cite{sarkar96}. A gainful analysis of the solvation
time correlation function into its partial components provided us insight into
the microscopic origin of this slow dynamics. We found that the major
contribution to the slow solvation of ions arose from their interactions with
the polar headgroups of the surfactants~\cite{bala1}.  
The observation of the slow  component in the decay of the
dipolar reorientational time correlation function of the interfacial water
molecules~\cite{bala2,paljcp} led us to investigate into the cause of the dramatic
slowing down of orientational relaxation. Careful analysis of the results demonstrated 
that it is caused by the long lived hydrogen bonds
that the water molecules form with the hydrophilic, and polar headgroups that
constitute the micellar surface~\cite{balaprl}. Specifically, we calculated the
lifetimes of such hydrogen bonds through appropriate time correlation functions
(TCF), and found that they are nearly 5 to 10 times longer lived than the
hydrogen bonds that water molecules form between themselves. We showed that
this longer lifetime profoundly influences the intrinsic dynamics of water
molecules at the interface, in terms of their ability to rotate, to translate,
and their capability to solvate free ions. 

Furthermore, in a recent communication,
we have presented, for the first time, microscopic evidence for the presence of
three kinds of water molecules in the interfacial
layer~\cite{cs02,bala_condmat,pal_jpcb}. This classification was based on the
number of hydrogen bonds that an interfacial water molecule makes with the
polar headgroups of the surfactants. We found that about 80\% of the water
molecules in the interface are singly coordinated to the surfactants, and that
the rest of the interfacial water molecules are evenly divided into categories that
either form two hydrogen bonds with two {\it different surfactant molecules}, or
those that do not form hydrogen bonds with the macromolecule at all. The former
class of water molecules that form one or two hydrogen bonds with the
surfactants can thus be called {\em bound} and the latter as {\em free}. We
introduced a nomenclature to identify such species as IBW2 -- for water
molecules that form two hydrogen bonds with two different surfactants, IBW1 --
for water molecules that form only one such hydrogen bond, and IFW -- for water
molecules that do not form such water-PHG (w-PHG) hydrogen bonds. The ratio of IBW2:IBW1:IFW
was found to be distributed as 1.1:8:0.9. Because of the enhanced strength of
the w-PHG bond, IBW2 has a lower potential energy than IBW1.  Yet, it is not
found in abundance as it is disfavored entropically, due to its requirement of 
a constrained
environment, and due to the rarity of events in which two surfactants need to
be present around it at the proper orientation, and geometry.  Thus, we had
shown that the bound and free classification, has both energetic, and
structural basis.  Needless to state is the
fact, that the water molecules at the interface retain their nearly tetrahedral
coordination, similar to bulk water, and that their coordination shell gets
completed by the formation of hydrogen bonds with other water molecules in the
vicinity.

 It should be pointed out at this stage that this classification of bound and free
 applies to {\it interfacial} water and therefore, does not refer to bulk water. Indeed,
 the existence of such free (or quasi-free, if we need to distinguish from bulk
 water) species {\it at the interface} could have far reaching consequence in 
 the understanding of both macromolecular activity, like molecular recognition, and
 water dynamics.

In light of our earlier observations on the slow dynamics of interfacial
water~\cite{bala1}, it would be interesting to examine if the bound-free
paradigm~\cite{nandi_bagchi97,zewailfeature} can have a much more
robust dynamical basis as
well. This is required because the hydrogen bond lifetime analysis provides a slow down
by a factor of 10 at the maximum. Thus, although the slow down of this lifetime is
probably the fundamental process, it does not fully explain the reason for
the dramatic slow down, to the extent of more than two orders of magnitude
that we observed in simulations and others observed in experiments.
In this article, we explore this aspect of the interfacial water
molecules.  Details of their structure and energetics have been reported
earlier~\cite{cs02,bala_condmat,pal_jpcb}. Here we focus on their dynamics. We find
that the {\it long time slow decay observed in the dipolar reorientational TCF
of the interfacial water molecules can be directly attributed to the doubly
bonded bound water species}. We also discuss results on the lifetimes of these
species, and we find surprisingly that the singly bonded water species is the
longest lived among the three species of interfacial water molecules.

 An additional important aspect of this paper is an analysis of the intermolecular vibrational
frequencies of the interfacial water.
With the recent advent of nonlinear spectroscopic methods that use lasers in
the midinfrared region, the dynamics of hydrogen bonds between water molecules
has been studied experimentally. In such a technique, one studies the change in
the intramolecular O-H stretching frequency for water molecules in different
environments. Such studies have shed considerable light on the dynamics of the
hydration layer around alkali and halide ions in aqueous salt
solutions~\cite{bakker_science,bakker_prl}. We foresee such experiments to be
performed for the bound water molecules described here.  Unfortunately, we
cannot study the change in the intramolecular O-H frequency of the bound water
molecules, as our water model has a rigid geometry.  However, we provide
details on the vibrational spectrum of the bound water molecules by studying
the Fourier transform of the velocity auto correlation function of water
molecules. We find that both the librational
and bending frequencies undergo a significant {\it blue shift}, signalling the
enhanced rigidity in the structure of the surface water. This blue shift
in libration frequency is in good agreement with recent incoherent, inelastic neutron scattering
data on macromolecular solutions~\cite{ruffle}. These results 
are expected to be rather general and should be relevant to the
understanding of dynamics of water near any hydrophilic surface.

 The organization of the rest of the paper is as follows. In the next section we present 
the details of the systems and of MD simulations. Section III present
 results on hydrogen bond lifetime. Section IV contains results of our
 analysis of the frequencies of intermolecular vibrations. Section V concludes
 with discussion of the results.

\section{Details of Simulation} 
The surfactant in our simulations is cesium perfluorooctanoate (CsPFO), which
has a carboxylate head group, and a fluorinated tail. It forms oblate
ellipsoidal micelles in water~\cite{boden7993,iijima98}. We have studied one
micelle, an aggregate of 62 of these surfactants in a simulation box of 10,562
water molecules. An atomistic molecular dynamics simulation of this system at
300K was performed, whose details have been reported
earlier~\cite{bala1,currsci,balaprl}. Here we provide essential information on
the methods used for analysis of the data pertaining to dynamics of the bound
water molecules. The simulations were performed in the NVT ensemble with an
integration timestep of 4~fs, using the RESPA algorithm~\cite{tuckerman92} and
the PINY\_MD package~\cite{martyna_unpub}.  The model for water is the
SPC/E~\cite{berendsen87}, and the interaction potential used has been described
earlier~\cite{bala1,bala2}.  Coordinates of all the atoms were recorded every
12~fs, and the correlation functions presented here were calculated from such a
trajectory. In order to calculate the atomic velocity TCFs, we generated a 15~ps
long trajectory, where the coordinates and velocities of all the atoms in the
simulation were recorded every timestep, so as to accurately obtain  the
initial fast decay of the velocity TCFs.  The hydrogen bond lifetime
correlation functions were obtained from a trajectory of length 50~ps, with a
time resolution of 12~fs.  The long time behavior of some of these functions
were obtained from another trajectory of length 3.3~ns, with a time
resolution of 1~ps. 

An important determinant of the dynamics of water molecules is the
reorientation of its dipole vector that can be probed with NMR measurements.
We have calculated the dipole-dipole TCF, defined as,
\begin{equation}
\rm C_{\bmu} (t) = \rm \frac{\langle{\bmu}_{i}(t+\tau) \cdot {\bmu}_{i}(\tau)\rangle}{
\langle{\bmu}_{i}(\tau)
\cdot {\bmu}_{i}(\tau)\rangle}
\end{equation}
where ${\bmu}_{i}(t)$ is the dipole moment vector of $i$-th water molecule at
time t, and the angular brackets denote averaging over water molecules, as well
as over initial configurations, $\tau$.

Further, we have calculated the lifetime of the hydrogen bonds that the bound
water molecules form with the polar headgroups of the surfactant. These have
been characterized in terms of two time correlation functions, S$_{\rm HB}$(t) and
C$_{\rm HB}$(t). These TCFs can be defined using the functions h(t), and H(t) which
signify the presence or absence of the hydrogen bond at any time $t$, and are
given as~\cite{stillinger_acp,rapaport,ferrario,luzar_chandler,chandra,starr,berne,balaprl},
\begin{eqnarray}
\rm h(t) & = & 1, \rm  if~a~pair~of~atoms~are~bonded~at~time~t, \nonumber \\
   & = & 0,  \rm otherwise \nonumber \\
\rm H(t) & = & 1, \rm if~a~pair~of~atoms~are~continuously~bonded~\nonumber \\
 & &     \rm  between~time~0~and~time~t \nonumber \\
  & = & 0, \rm \nonumber otherwise
\end{eqnarray}

Using these definitions, the bond lifetime correlation functions are defined as,
\begin{eqnarray}
\rm S_{HB}(t) & = & \rm \frac{\langle h(\tau)H(t+\tau)\rangle}{\langle h \rangle} \nonumber \\
\rm C_{HB}(t) & = & \rm \frac{\langle h(\tau)h(t+\tau)\rangle}{\langle h \rangle} 
\end{eqnarray}
S$_{\rm HB}$(t) probes the continuous existence of a hydrogen bond, while
C$_{\rm HB}$(t)  allows for the reformation of a bond that is broken at some
intermediate time. The former is thus a strict definition of the hydrogen bond
lifetime, while the latter is more permissive. The true lifetime of a hydrogen
bond lies somewhere in between the two time constants obtained from these
functions.

In defining the presence of a hydrogen bond, we employ two distance conditions
and one energy criterion, as advocated earlier~\cite{starr}.
These criteria and their legitimacy with regard to the hydrogen bond formed
between water and the polar head group have been discussed
earlier~\cite{cs02,bala_condmat,pal_jpcb}. Here we just state the definition.
We assume a water molecule to be hydrogen bonded to a polar headgroup, if (a)
the distance between the oxygen of the water molecule and the carbon of the
headgroup is within 4.35\AA, and (b) the distance between the oxygen atom of
the water molecule and any of the oxygens of the polar headgroup is within
3.5\AA, and (c) the pair interaction energy between the water molecule and the
headgroup is less than $-6.25$~kcal/mole.

We have also calculated the lifetime correlation function of the interfacial
water species, denoted as S$_{\rm W}$(t), and C$_{\rm W}$(t). Their defnitions are similar
to S$_{\rm HB}$(t)  and C$_{\rm HB}$(t), above. However, they are dependent on
functions, $\rm h(t)$ and $\rm H(t)$ that are unity when a particular water molecule is
of a certain type, and are zero, if it is not.  Thus, these need to be
distinguished from the S$_{\rm HB}$(t) and C$_{\rm HB}$(t) functions which are
descriptions of the lifetime of the hydrogen bond.  In making this distinction,
we allow for the existence of a water molecule in any of the three states --
IFW, IBW1, or IBW2, despite the ephemeral loss of a hydrogen bond with a
particular PHG. 

 In order to study the  vibrational spectrum of the intermolecular hydrogen
 bonds, especially those of the bound water at the interface, we have
 carried out the Fourier transform of the velocity auto correlation function of water
molecules defined as, $\vec{v}$
\begin{equation}
\rm C_{vv}(t) = \rm \frac{\langle \vec{v}_i(t+\tau)\cdot\vec{v}_i(\tau)\rangle}
{\langle \vec{v}_i(\tau)\cdot \vec{v}_i(\tau)\rangle}
\end{equation}
where ${\bf v}_i(t)$ is the velocity vector of atom $i$ in a water molecule at time
$t$, and the angular brackets denote averaging over all atoms of such type in
the system, and over initial times, $\tau$. The Fourier transform of the velocity
correlation function provide a direct measure of the vibrational density of states as
a function of the frequency.
%*******************************************
\section{Results : Lifetime of Hydrogen Bond and of Interfacial Species}

We have provided a schematic illustration in Figure~1 of the two bound interfacial
water species, that describes their bonding pattern with the headgroups. The
free water molecules are present in the interfacial region, however they are
not oriented properly to form a hydrogen bond with the PHG. As discussed
earlier~\cite{cs02,bala_condmat}, the concentration of IFW:IBW1:IBW2
is 9:80:11.

\subsection{Lifetime of water-surfactant hydrogen bond}
We had shown earlier that the hydrogen bond formed between the water molecules
at the interface and the polar headgroups of the surfactant has a much larger
lifetime than the hydrogen bonds that water molecules form between
themselves~\cite{balaprl}.  In Figure~2, we show similar lifetime correlation
functions for the individual species in the interfacial region.  Consistent
with our earlier observations~\cite{balaprl}, the S$_{\rm HB}$(t) functions for
both the bound species are much slower than the corresponding function for the
water-water hydrogen bond in pure water.  The function for IBW1 water is slower
than that for the doubly hydrogen bonded IBW2 water species. The latter is in a
constrained bonding arrangement~\cite{cs02,bala_condmat}, and thus the
average lifetime of any one of its hydrogen bond is shorter than that formed by
the IBW1 water species. In our earlier work~\cite{balaprl}, the
condition on the pair energy of the water molecule with the PHGs, to determine
the presence of a hydrogen bond was not included. The current results, which
include such an energy criterion, confirm our findings on the long lifetime of
the bound water-PHG hydrogen bond~\cite{balaprl}. In addition, we are also able
to clearly distinguish the contributions from the two types of interfacial
bound water.

The C$_{\rm HB}$(t) function, presented in the inset to Figure~2, shows a similar trend as
the S$_{\rm HB}$(t) function. As noted earlier, this function allows for the
reformation of the hydrogen bond, and thus would take into account, recrossing
of the barrier, as well as long time diffusive behavior.  Again, we find a
characteristic slow decay in its relaxation. The relatively longer lived
C$_{\rm HB}$(t) function for the IBW2 species could come from the break and
reformation of the bond over short distances or angles. The IBW2 species will have a higher
propensity to reform hydrogen bonds than the IBW1 species, as the latter is
only singly hydrogen bonded to the PHGs.  Note that in this formalism, we only
track the existence of a particular bond of the interfacial water species, and
hence the behavior of the functions for IBW1 and IBW2 may not be very
different.

\subsection{Lifetime of interfacial water species}
A bound water species could remain bound (to another PHG) even if the hydrogen
bond it had originally formed with a particular PHG is broken.  Hence, it is
important to obtain information on the lifetime of the species, apart from the
lifetime of the w-PHG bond discussed in Figure~2.  We present this data on the
lifetime correlation function of the species in Figure~3.  As discussed in the
previous section, these functions are essentially similar to the S$_{\rm HB}$(t)
functions defined above, but instead are defined for the identity of the
species involved.  We observe that the IFW species is short lived, and that the
IBW1 species is the longest lived. In our earlier work, we had concluded that
the IBW1 species is the thermodynamically stable
species~\cite{cs02,bala_condmat}. The current calculations indicate that
this species is also dynamically more stable than the IBW2 or the IFW species.
The IBW2 species, although preferred on energetic grounds due to the
contribution from two strong w-PHG hydrogen bonds, is disfavored entropically. It
requires the simultaneous presence of two surfactant headgroups in the proper
geometry, which is rare.  We present the C$_{\rm W}$(t) functions in the inset to
Figure~3. Again, the function for IBW1 is much slower than that of IBW2 or 
that of IFW.  The slow component presumably arises out of a few water 
molecules (say, of IBW1 type) that lose their identity and reform a w-PHG 
hydrogen bond either with the same PHG or with another PHG. These molecules 
probably have long residence times within the first few hydration layers around
the micelle.

Within the interfacial region, there is a constant exchange of water molecules
between the three states, IBW2, IBW1, and IFW. The microscopic reactions
between IBW2 and IBW1 on the one hand, and IBW1 and IFW on the other, are
reversible, and are described by four distinct rate constants, as described
below. 
\begin{equation}
 \rm IBW2 \rightleftarrows^{k_{21}}_{k_{12}} IBW1\rightleftarrows^{k_{1F}}_{k_{F1}} IFW
\end{equation}
Determining these rates in such a complex system that is also {\em open} to
bulk water, is a challenge to us and to other simulators, although the basic
formalism exists~\cite{chandler_rate}. Here, we concentrate on generic features
that one can derive based on the concentrations of these species. The reactions
delineated here are the elementary events, and hence the condition of detailed
balance must be obeyed for each of these reactions. The ratio of IBW1 to IBW2
is approximately 8. Hence the rate of production of the IBW1 species from the
IBW2 species ($\rm k_{21}$) must be a factor of eight larger than the rate of
the reverse reaction. This rationalizes the shorter lifespan of the IBW2
species. Note that this ratio of eight in the rates of interconversion between
IBW1 and IBW2 species will be reduced by the other reaction proceeding in the 
system, that between IFW and IBW1. This too will add to the attrition in the 
lifetime of the IBW1 species.
Fitting the S$_{\rm HB}$ data of lifetimes presented in Figure~3 to multiexponential forms, we
find that the average lifetime of the IBW2 species is 0.62~ps, and that of the IBW1
species is 2.26~ps. Thus the IBW1 species is around 3.6~times longer lived than the IBW2
species, which is not inconsistent with the arguments presented above. 

\subsection{Reorientational dynamics of interfacial water}
Having determined the intrinsic lifetimes of these interfacial water species,
and of the strong hydrogen bonds that they form with the polar headgroups of
the surfactant molecules that constitute the micellar surface, we turn our
attention now to their single particle dynamics, namely their ability to
reorient.  The behavior of the dipolar reorientational TCF of the individual
bound water species illuminates our understanding on the important issue of the
slow decay observed in our simulations
earlier~\cite{bala1,currsci,berkowitz02}. With this objective, we have
calculated the dipole-dipole TCFs of the IFW, IBW1, and IBW2 species, which are
compared to the function for water molecules far away from the micelle in
Figure~4.  These data were obtained with a time resolution of 12~fs. The data
for water molecules that are at least 25\AA~away from the micellar surface shows
that they behave just like bulk water molecules. The inset to the figure shows
the very fast decay of the curve corresponding to the IFW species, very similar
to that of water far from the micelle.  This validates our christening this
species as free interfacial water, IFW.  The plot also displays the
corresponding TCF for the IBW1 and IBW2 species. The reorientation of the IBW1
water, possessing a single, strong hydrogen bond with the PHG is slowed down
relative to that of the IFW species, while the IBW2 water species, being in a
constrained environment is able to rotate only within a narrow cone. This is
borne out by the plateauing of the dipolar TCF after a few picoseconds, found
in the data for the IBW2 species. The strict definitions of a water-PHG
hydrogen bond precludes us from probing the TCFs any further than the durations
shown in the figure, as these species themselves cease to survive, a few picoseconds
beyond the maximum times for which the TCFs have been displayed.  However, a
bound water is unlikely to lose its memory of orientation, once it has
``broken'' its w-PHG hydrogen bond. It is more likely that its orientation is
preserved for some longer duration, and that the strict definition of the
existence of the bound nature of water prevents one from probing the
orientation any further.  This indicates that we could continue to probe the dipolar TCFs of
these waters for a longer duration, if we relax the conditions on the nature of
the bound water species. We describe such a calculation below.

Water molecules within a 10\AA~hydration layer around the micelle exhibit a
pronounced slowing down in their reorientation~\cite{bala1,berkowitz02}. The
long time behavior of the C$\mu$(t) shows a near-constant value, indicating
that the molecules are not able to explore fully all possible orientations. In
our simulations, the near-constant value of C$\mu$(t) extends to a few hundreds
of picoseconds and appears to be a general feature to interfacial water near
hydrophilic surfaces in complex systems.  This observation may seem apparently
inconsistent with the data presented in Figure~3 or Figure~4, where the bound water of
either the IBW1 or IBW2 type, are seen to lose their identity within few tens
of picoseconds. The resolution of this apparent conflict, lies, as we discussed
earlier, in the rather strict definition of the lifetime of the bound species.
We have calculated the same functions exhibited in Figure~4, with a coarser
resolution of 1~ps, and display them for the bound water species in Figure~5.
Shown here are the C$_\mu$(t) functions obtained from a trajectory where the
coordinates of all the atoms were stored every 1~ps, instead of the 12~fs used
in Figure~4.  This effectively allows for the consideration of reformation of
the w-PHG hydrogen bond, thus extending the lifetime of the bound water species.
We are thus able to observe the reorientational TCFs of these species for a
much longer duration than what was possible in Figure~4. Such a data, presented
in Figure~5 clearly shows the existence of very slow dynamics in the timescales of hundreds
of picoseconds. More importantly, the reorientational TCF obtained for all bound water species,
irrespective of the IBW1 or the IBW2 type, exhibits a plateau followed by a
slow decay. 
These results shows that exchanges between the IBW1 and the IBW2 states of water molecules
artificially prevents one from obtaining the dipolar reorientational TCF of either of these
individual species for times longer than their ``intrinsic'' lifetimes. However, the TCF for all
bound water species, irrespective of their type, clearly proves the existence of a 
long-lived plateau running probably into hundreds of picoseconds. This is the first, direct
microscopic evidence for such a slow relaxation arising out of water species bound to 
a hydrophilic surface. As mentioned earlier, these results should be of relevance to
other hydrophilic surfaces as well.

It is thus clear that the long-time slow decay in the dipolar
reorientational TCF arises out of \\
(i) the orientational locking-in of the IBW2 species, due to its two strong w-PHG hydrogen bonds\\
(ii) the interconversion between the IBW1 and IBW2 species.

The latter, a secondary process, merits further discussion. From the TCFs shown
in Figure~5, one can obtain average time constants for the relaxation of the 
reorientation of the dipole. This is certainly valid if a water molecule, for example, was of
the IBW1 type at time zero, and remained so thereafter. However, the average
lifetime of the w-PHG hydrogen bond is of the order of 2 to 40~ps (see
Figure~2), and hence one can expect a fair number of water molecules to
interconvert between the IFW, IBW1, and IBW2 states within the interfacial
layer. Hence the said IBW1 water molecule may form an additional w-PHG bond and
change its nature to that of a IBW2 type. This would, in effect, set the clock
for reorientation of its dipole to the dipolar TCF corresponding to the IBW2
type shown in Figure~5, which is much slower than that for the IBW1 species.
Such interconversions between the interfacial water species will add to the
slow decay of the intrinsic species shown in Figure~5 and would contribute,
specifically, to the intermediate time scales (tens of picoseconds) in the TCF, while the
orientational locking-in of the IBW2 species will contribute to the long time
slow decay (hundreds of picoseconds). Further, a few members of the IBW2 type can be long lived, and they
too would contribute to the long time behavior of the total dipolar
reorientational TCF.  It is the sum effect of these processes that we had
observed earlier in our preliminary investigations of the reorientations of
interfacial water molecules. This interpretation of the exchange between the
species present at the interface to contribute to the slow dynamics is
consistent with the postulates of Nandi and
Bagchi~\cite{nandi_bagchi97,zewailfeature}.

%*************************************************************************
\section{Vibrational Spectrum of Interfacial Water and Density of States}

 Thus far we have provided microscopic evidence for the slowdown in the
dynamics of interfacial water due to the presence of the polar, fluctuating
micellar surface.  The interface influences not only single particle
quantities such as the dipolar reorientation of the water molecules, but it
also affects the ability of water molecules to librate (hindered rotation), and
their ability to alter the hydrogen bond network through bond bending
excitations. We probe such molecular motions through calculations of the
velocity time correlation function of the water molecules and the density of
states of the respective motions.

Low frequency Raman and neutron scattering experiments on aqueous protein
solutions have  shown the presence of {\it excess density of states}, called the
{\it boson peak},  at around 3~meV (24~cm$^{-1}$)~\cite{boson_expt} 
which has been corroborated by molecular dynamics
simulations~\cite{tarek_tobias,rocchi,canni_boson}. This boson peak is due
to low frequency collective vibrational mode of the biomolecule. 
In this work, however, we concentrate
on excitations of relatively higher frequencies. Specifically, we study
the effect of the interface on the O..O..O bending, O..O stretching, and the
librational modes of water, that are observed at around 50~cm$^{-1}$, 200~cm$^{-1}$,
and at 500~cm$^{-1}$~\cite{ohmine,vibr_expts}, respectively, in bulk water.
Note that all the three are inter-molecular vibrational modes and involve
hydrogen bonding. All these modes play an important role in the dynamics of
chemical processes in water, such as solvation dynamics and charge
transfer~\cite{roybagchi,biswasbagchi}.

 We have obtained the velocity autocorrelation functions of both
the oxygen  and the hydrogen atoms of interfacial water molecules, and
have obtained their power spectra. The results indicate strong effects
of the micellar surface.

The computed velocity correlation functions are presented in Figure~6 where we have also
compared them with corresponding functions for water molecules far away from
the micelle. In Figure~6a, we present the C$_{\rm vv}^{\rm O}$(t) for the oxygen atoms
of the interfacial water species. As expected, there is a clear trend in the
variation of this function, particularly near the first minimum that
corresponds to backscattering of the atoms. The minimum for IBW2 is deeper than
that for the rest of the species, a direct consequence of the enhanced rigidity
of its environment. The function for water molecules far away from the micellar
surface is identical to its behavior in bulk.  The function for IFW follows
closely that of bulk water, however it exhibit a shallower first minimum than
bulk water. We had shown earlier~\cite{pal_jpcb} that the first coordination 
shell (up to a distance of 3.5\AA) around the IFW species contains
4.4 water molecules and 0.3 PHG oxygens, while that around the IBW1 species contains 4.1
 water molecules and
1.2 PHG oxygens, and that of the IBW2 species contains 3.0 water molecules and 2.1 PHG oxygens. 
This marginal decrease in the coordination number around the IFW species could be 
a prime reason for the shallower first minimum in its C$_{\rm vv}^{\rm O}$(t) than that 
for bulk water.
These broad conclusions are substantiated by an
analysis of the C$_{\rm vv}^{\rm H}$(t) function for the hydrogen atoms of the
interfacial water species presented in Figure~6b. 
The enhanced rigidity around
IBW2 water manifests itself in a deeper minimum and again, the data for IFW
species mirrors closely the behavior for water in bulk. However the value for
the IFW species at the first minimum is less negative than that for bulk water. Lee 
and Rossky~\cite{lee_rossky} had found a similar behavior for water 
molecules present near a hydrophobic surface. This indicates that the IFW species  
might be somewhat buried within the micelle, in proximity to the CF$_2$ groups 
adjacent to the polar headgroup.  The data for IBW1, the
abundant species in the interface lies between the functions for IFW and that
of IBW2.

The power spectra, obtained by a Fourier cosine transform of these velocity
auto time correlation functions, are displayed in Figure~7. In Figure~7a, we
provide this data obtained from oxygen atoms, for all interfacial water
species, i.e., for water molecules that reside within a distance of 4.5\AA~
from any of the surfactant headgroups.  Transforms of individual water species
were noisy, and hence are not provided here.  Relative to water in the bulk, we
observe a clear blue shift in the frequency corresponding to  the O..O..O
bending mode by about 40~cm$^{-1}$  for these interfacial water molecules while
the stretching mode seems not to be affected by the interface, consistent with 
earlier observations on aqueous protein solutions~\cite{steinhauser_96,rocchi}.
The most significant change in the
vibrational spectrum occurs for the librational mode of the water molecules
which can be studied from the power spectrum for the hydrogen atoms of the
interfacial water species.  Again, the trend observed in Figure~6b is seen in
their power spectra.  The librational mode of IBW2 is most affected, and is
shifted considerably, by about 150~cm$^{-1}$, whereas that of the IBW1 species
is blue shifted by about 100~cm$^{-1}$ relative to water bulk and is observed
at 600~cm$^{-1}$.  {\it These data agree well with recent incoherent, inelastic
neutron scattering experiments on aqueous DNA and membrane solutions,
where the vibrational signature of interfacial water molecules was found to be
predominant in the 400-600~cm$^{-1}$ range}~\cite{ruffle}. The hydrogen atoms
also contribute to the O..O..O bending mode  at around 60~cm$^{-1}$, whose
shift to higher frequencies is also observed from the oxygen spectrum. 
These frequency shifts in the bending and librational modes must arise
from the strong, and longer lived w-PHG hydrogen bond. Evidence for the
strength, and the lifetime of these bonds have been provided earlier.
Therefore, one may conclude that water structure on the surface is more rigid.
While this is of course expected on simple arguments, the strong effect on
bending and librations, while the negligible effect on stretching were not
anticipated, but is in agreement with IINS results. This needs further
study.

%*********************************************
\section{Conclusions}

The results of the present atomistic MD simulations provide microscopic
explanation for several of the observations reported
earlier~\cite{bala2,paljcp,currsci,balaprl}.  We find 
that the hydrogen bonds between the bound water molecules and the
polar headgroups are much longer lived, on the average, than the hydrogen bonds that
water molecules form among themselves.  The lifetime of the w-PHG hydrogen
bond of the IBW2 type of water molecule is shorter than that of the IBW1 type.
This can be rationalized in terms of the constrained nature of the environment
around the IBW2 water molecule. This aspect is
also reflected in the lifetimes of the intrinsic species. The IBW1 water is
dynamically more stable than the IBW2 species, which is consistent with our
observations on its concentration, energetics and
environment.

We have also calculated {\it individual} dipolar
reorientational time correlation functions for the three species of interfacial
water molecules, and have compared them to the corresponding function for water molecules in
bulk. The IFW water molecule is able to reorient in the same timescale as the 
bulk water. {\em Thus, the interfacial free water appears to be energetically,
structurally, and dynamically rather similar to, albeit a bit slower than,
the water in the bulk}. The bound water molecules exhibit very different 
dynamical properties. The IBW1 water
molecule exhibits a slow relaxation with the longest component around several
tens of picoseconds, while the IBW2 water molecule exhibits a long lived
plateau region in the time correlation function. This can be ascribed to the
two strong w-PHG hydrogen bonds that this species makes, resulting in a
significant loss  of orientational freedom. About 10\% of the interfacial water
molecules are of the IBW2 type, and hence this plateau can be directly
implicated in the long-time relaxation of the dipolar TCF of all the
interfacial water molecules that we had reported earlier~\cite{bala2,currsci}.
The current set of results thus explain, comprehensively, the microscopic
origin of the slow decay in the reorientational TCF of interfacial water
molecules in aqueous macromolecular solutions.

A significant finding of the present work is the modification in
the vibrational spectral
features of the interfacial water molecules. 
We observe a  blue shift in the
O..O..O bending mode of about 40~cm$^{-1}$. The librational mode of the
water molecule at around 500~cm$^{-1}$ undergoes a blue shift of about
100~$cm^{-1}$. Surprisingly, the 200~cm$^{-1}$
translational, O..O stretching mode is entirely 
unaffected due to the interface. These results are somewhat surprising and
indicate that the the potential surface around the interfacial water molecules
is more rigid along the orientational degrees of freedom rather than along the
translational degrees. This conclusion is also consistent with observations 
on the relative effects of the interface on translational diffusion of water
molecules and their orientational relaxation~\cite{paljcp,berkowitz_jcp03}.
Our results on the vibrational spectra also adds support to recent IINS data on
aqueous membrane and DNA solutions, where the interfacial water molecules have
an altered vibrational spectrum in the 400-600~cm$^{-1}$ range~\cite{ruffle}.
We plan to probe these aspects further by a normal mode analysis of the
potential energy hypersurface. This could explain the reason for the lack
of change in the 200~cm$^{-1}$ intermolecular vibrational mode. This point
deserves further study.

\section{Acknowledgements}
We thank Dr. Francis Starr for pointing out Ref.~\cite{stillinger_acp}. SB and
BB acknowledge financial support from the Council of Scientific and Industrial
Research, and the Department of Science \& Technology, India.

\newpage
\centerline{\bf FIGURE CAPTIONS}
 
\noindent{\bf Figure 1} Schematic representation of IBW1 and IBW2 types of interfacial
water molecules. Water molecules at the interface that are not 
hydrogen bonded to any surfactant, but are instead coordinated fully to other water
molecules in the vicinity are denoted as IFW.

\noindent{\bf Figure 2} Hydrogen bond lifetime correlation function, 
S$_{\rm HB}$(t) for IBW1 (continuous line) and for IBW2 (dashed line) 
types of 
water molecules. Inset show the decay of C$_{\rm HB}$(t) for the same.

\noindent{\bf Figure 3} Species life time correlation function, S$_{\rm W}$(t) for the
different interfacial water molecules. Inset show the C$_{\rm W}$(t) function 
for the same. Numerical data points are shown infrequently for clarity.

\noindent{\bf Figure 4} Dipolar reorientational time correlation function, C$_\mu$(t), 
for the interfacial water molecules. The inset compares the data for the
IFW water species and for water molecules that are at least 25\AA~away from the
micellar surface.  The latter behave like water in bulk.
The time resolution of the 
underlying trajectory used in this calculation is 12~fs.
Numerical data points are shown infrequently for clarity.

\noindent{\bf Figure 5} Dipolar reorientational time correlation function, C$_\mu$(t), for
bound water molecules. The data denoted by the legend, IBW1-IBW2, corresponds to all 
bound water molecules irrespective of their being in state IBW1 or IBW2.  The time resolution of the 
underlying trajectory used in this calculation is 1~ps.
Numerical data points are shown infrequently for clarity.

\noindent{\bf Figure 6a} Normalized velocity auto correlation function,  C$_{\rm vv}^{\rm O}$(t), of the 
oxygen atoms for various interfacial water molecule types compared with that for 
water molecules that are at least 25\AA~away from the
micellar surface.  The latter behave like water in bulk.
Numerical data points are shown infrequently for clarity.

\noindent{\bf Figure 6b} Normalized velocity auto correlation function, C$_{\rm vv}^{\rm H}$(t), of the 
hydrogen atoms for various interfacial water molecule types compared with that for 
water molecules that are at least 25\AA~away from the
micellar surface.  The latter behave like water in bulk.
Numerical data points are shown infrequently for clarity.

\noindent{\bf Figure 7a} Power spectrum of the C$_{\rm vv}^{\rm O}$(t) function, for 
oxygen atoms of water molecules
lying within 4.5\AA from any polar head group of the micelle compared with that for 
water molecules that are at least 25\AA~away from the
micellar surface.  The latter behave like water in bulk.

\noindent{\bf Figure 7b} Power spectrum of the C$_{\rm vv}^{\rm H}$(t) function, for 
hydrogen atoms of various interfacial water molecule types compared with that
for water molecules that are at least 25\AA~away from the
micellar surface.  The latter behave like water in bulk.

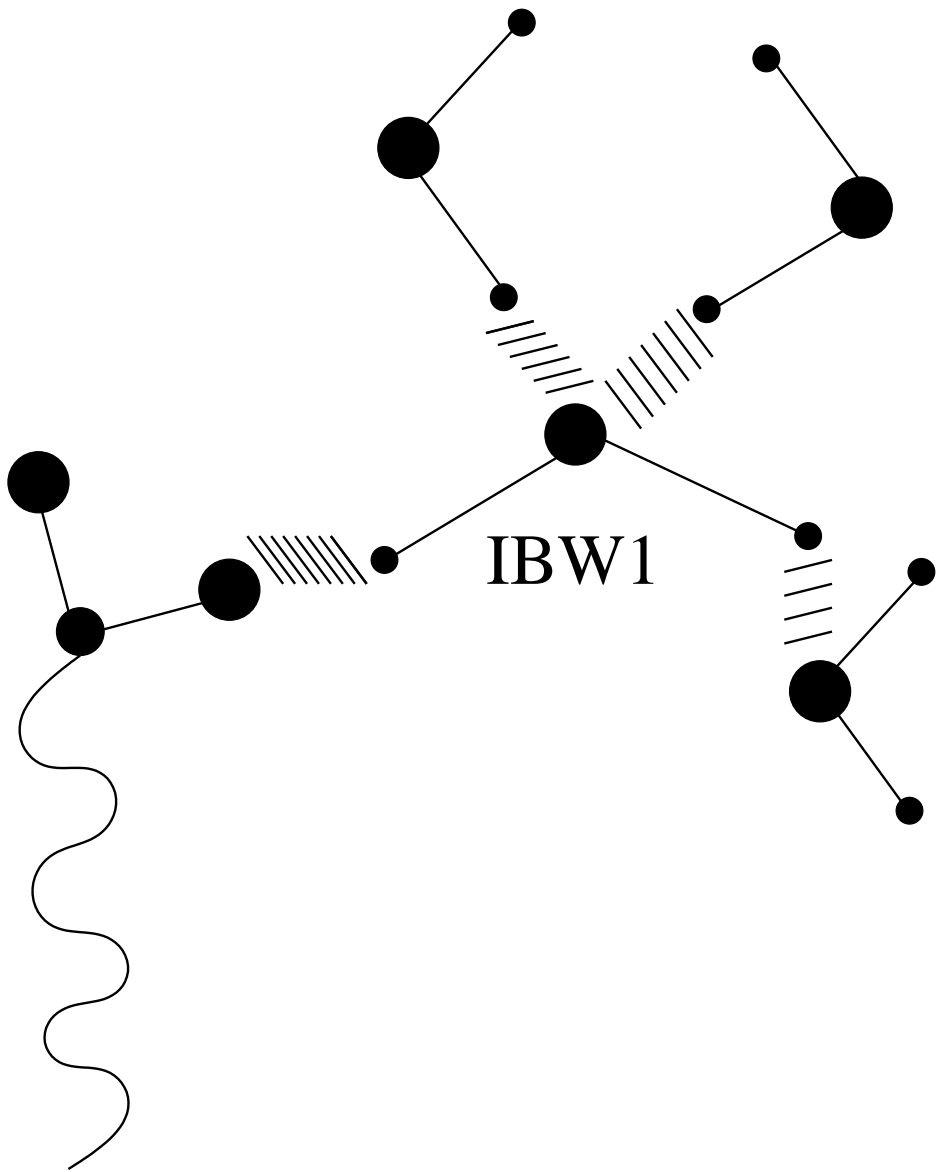
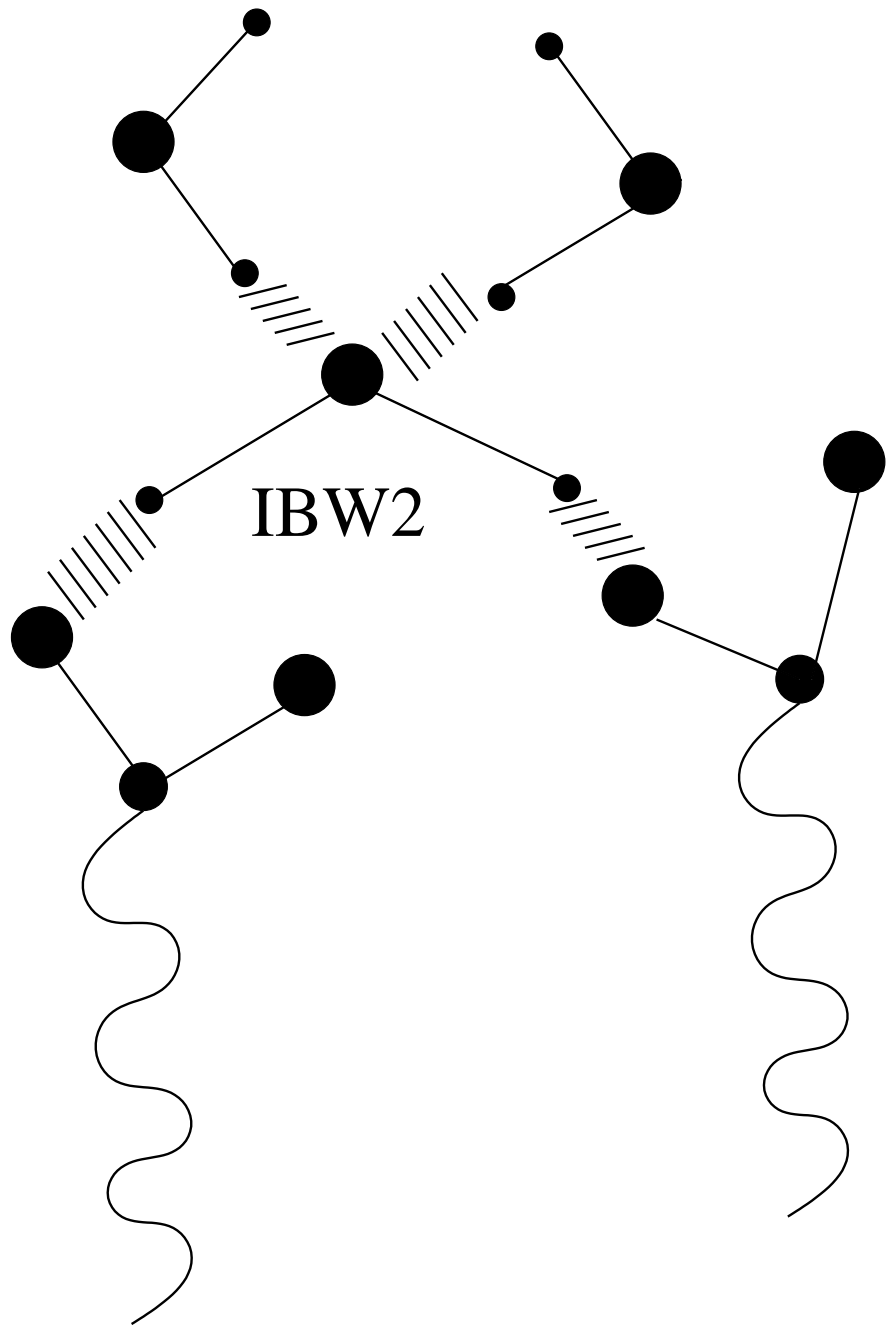
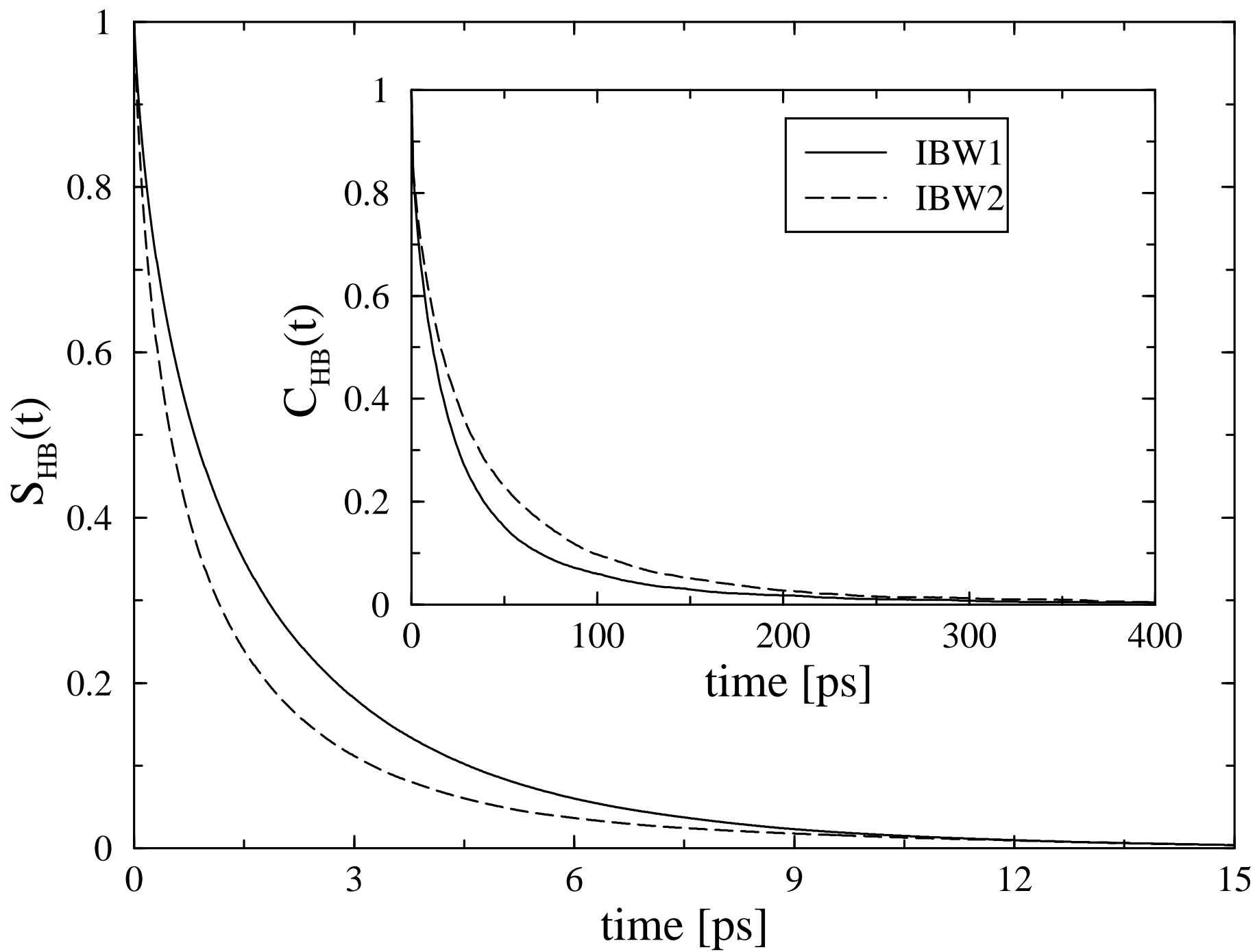
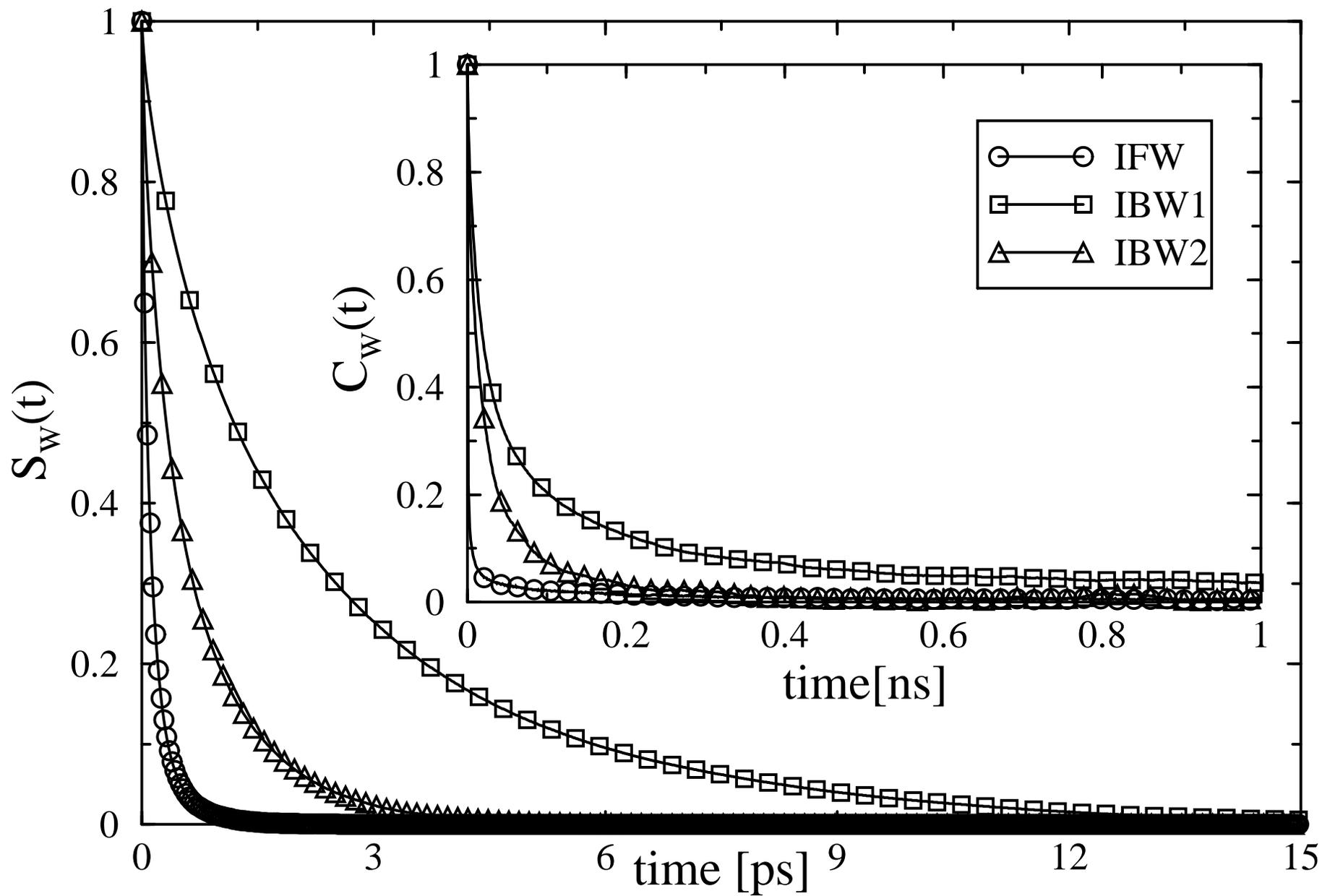
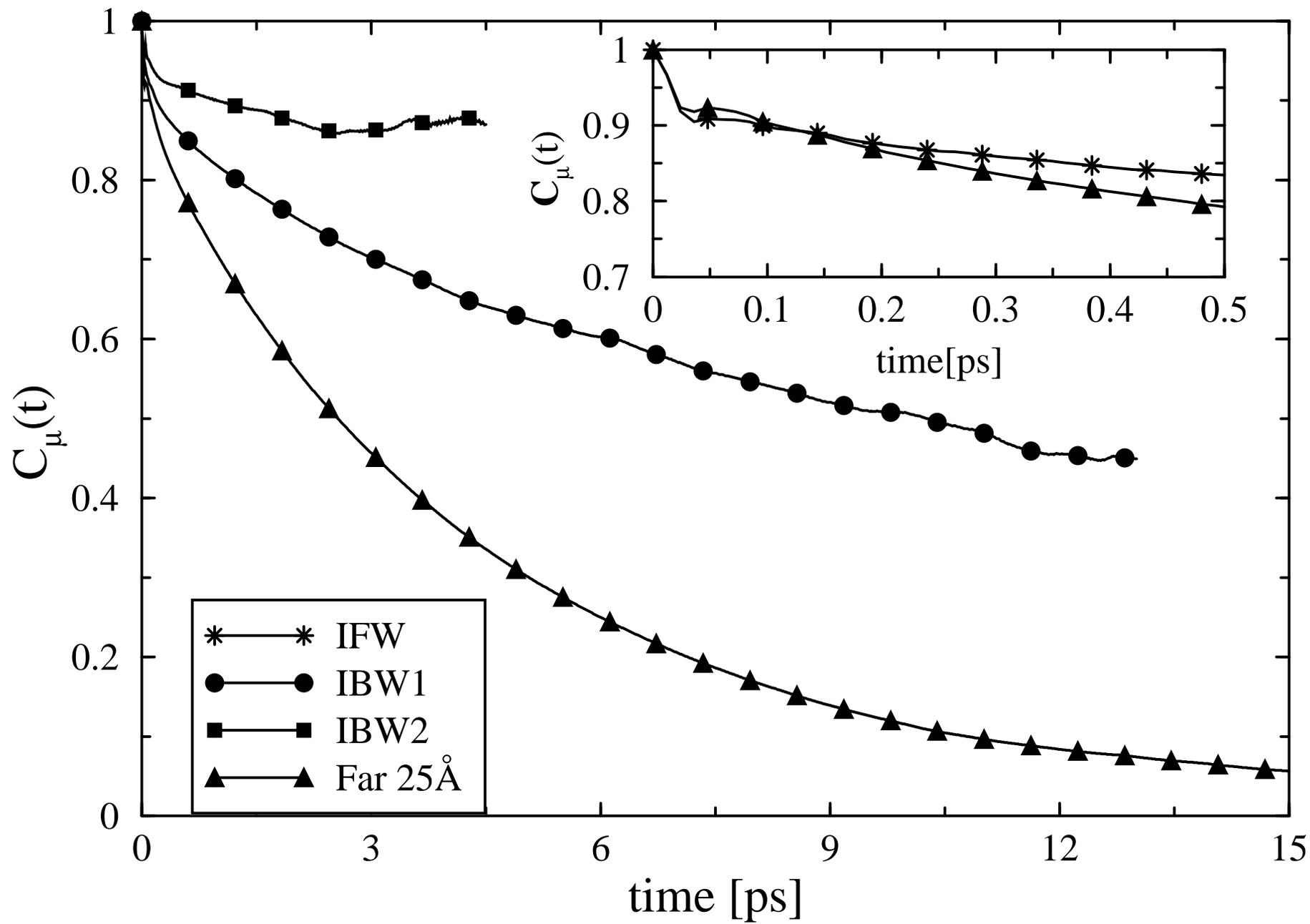
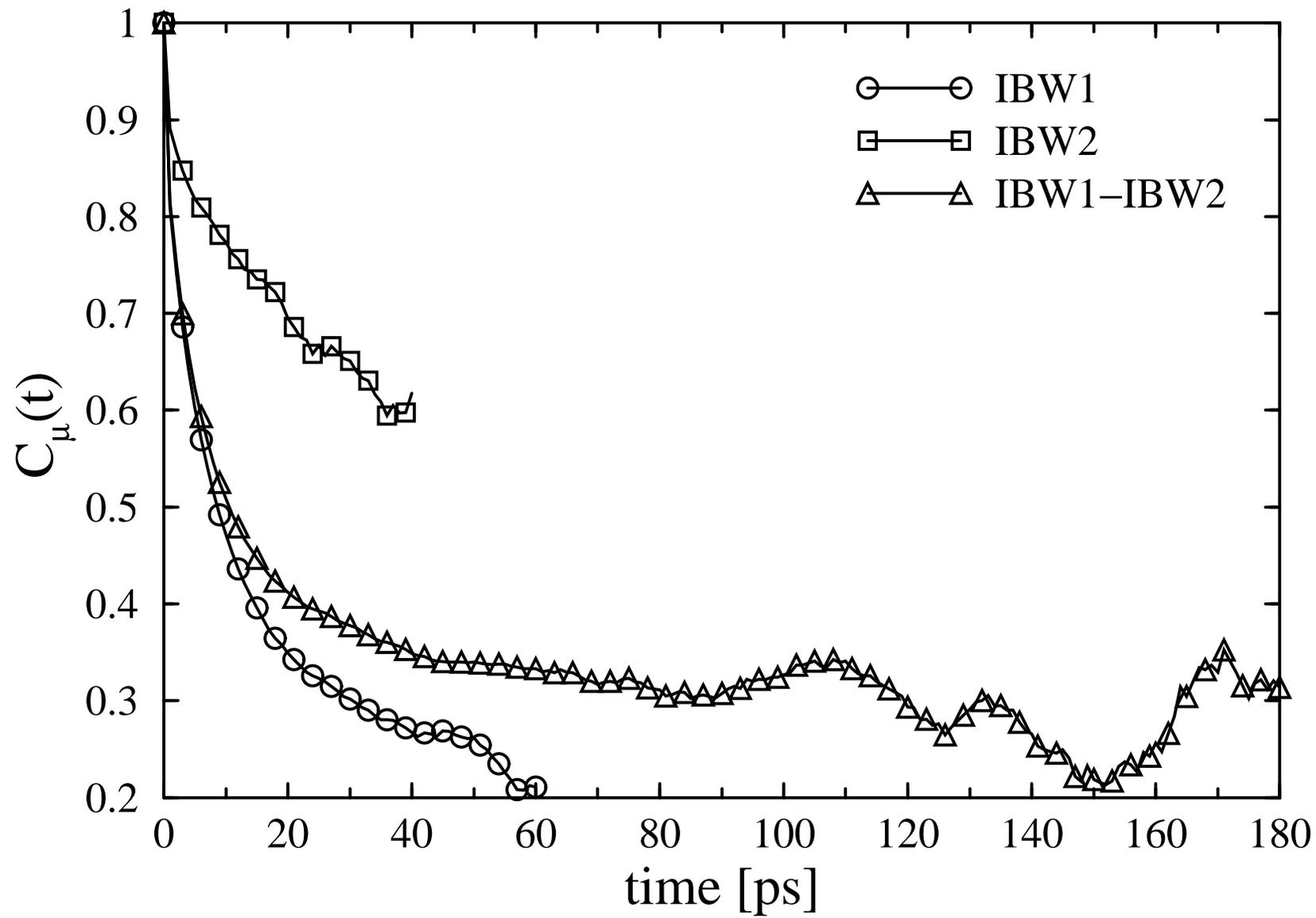
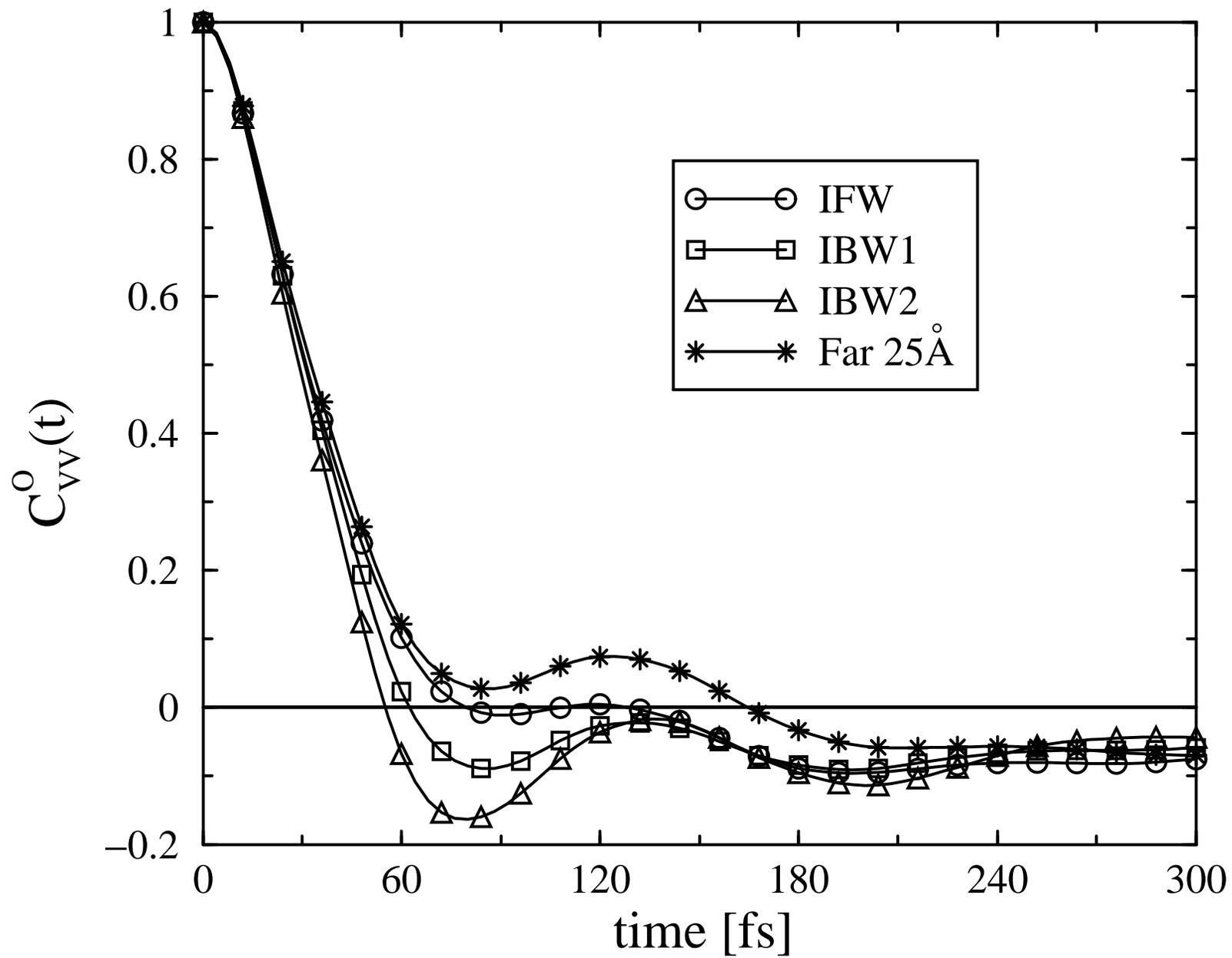
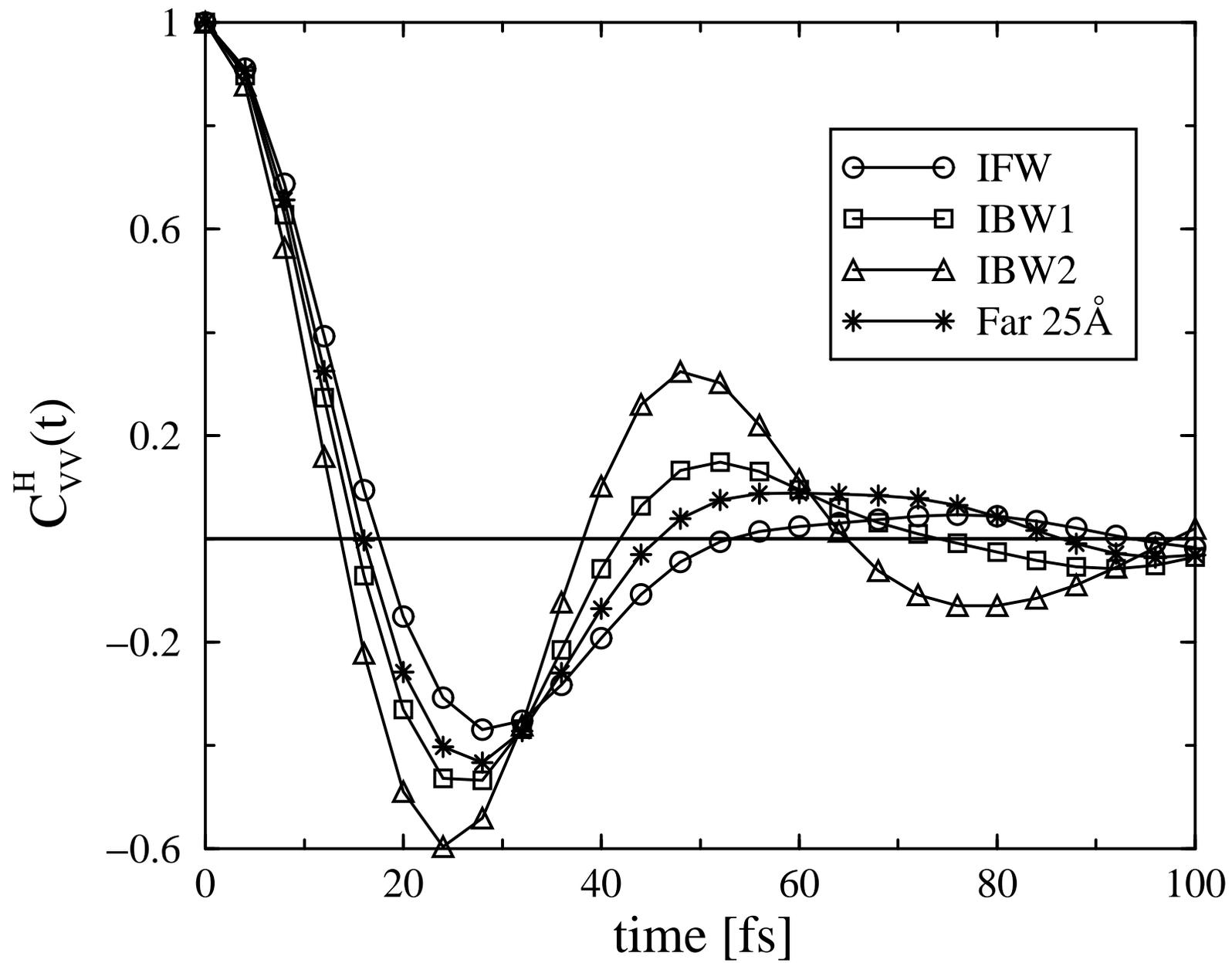
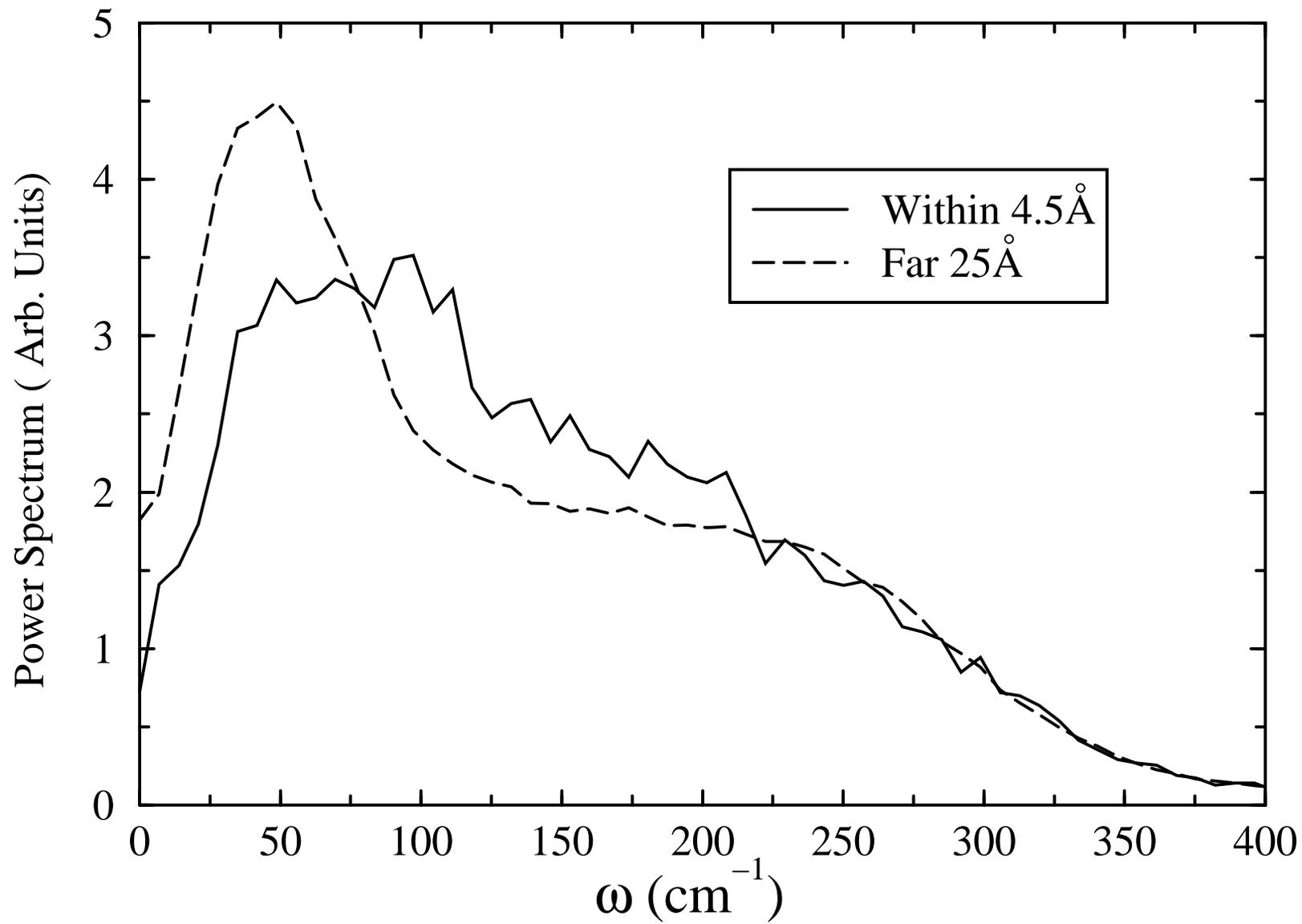
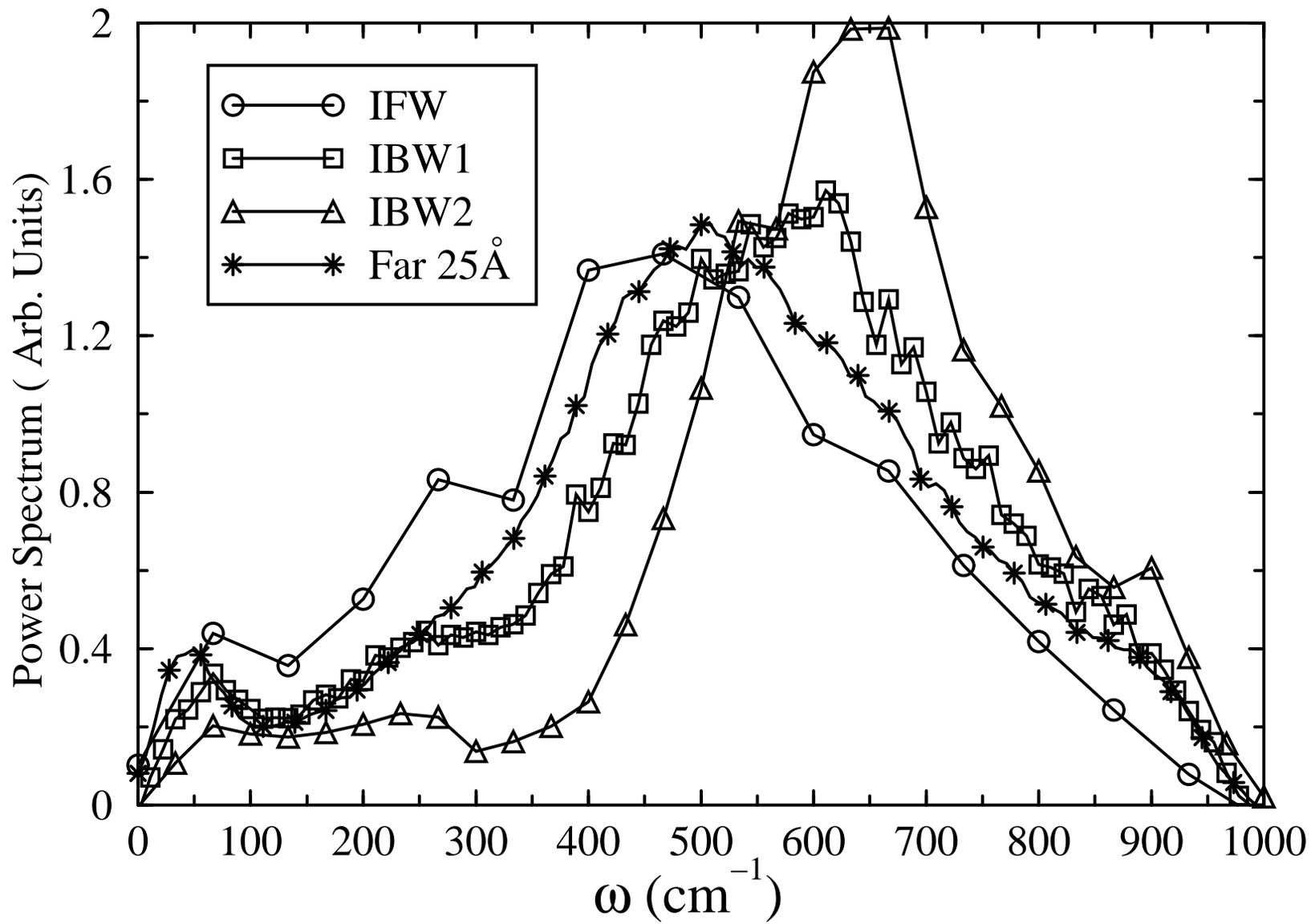
\end{document}